\documentstyle[twocolumn,epsf,aps]{revtex}

\begin{document}
\title{\bf\Large
Conductance through a Magnetic Domain Wall in Double Exchange System
}

\author{
Masanori {\sc Yamanaka}
\cite{email1}
and
Naoto {\sc Nagaosa}
\cite{email2}
} 

\address{
Department of Applied Physics, University of Tokyo,
Bunkyo-ku, Tokyo 113 \\
}

\date{\today}

\maketitle
\begin{abstract}
The conductance through a magnetic domain wall  is calculated for
the double exchange system as a function of energy
and the width of the domain wall.
It is shown that when the carrier density is low enough, 
the blockade is almost complete even for the 
smoothly varying spin configuration, i.e., large width of the domain wall.
This result is applied to the manganese oxides.
\end{abstract}

\vspace{3mm}
{[{\bf\small KEYWORD}:
double exchange system,
doped Mott insulator,
La$_{1-x}$Sr$_x$MnO$_3$]
}

\vspace{10mm}


Recently intensive studies have been focused on the colossal
magnetoresistance (CMR) in double exchange systems, e.g., manganese oxide
La$_{1-x}$Sr$_x$MnO$_3$.\cite{REFchahara,REFhelmolt,REFtokura,REFjin} 
The conduction electrons in $e_{\rm g}$ orbitals are moving in the background 
of the $t_{\rm 2g}$ spin configuration,
and the basic mechanisms of this phenomenon is that
the coherent transfer of electrons is enhanced 
when the external magnetic field forces the spins to align.
In these materials, the $t_{\rm 2g}$ spin is localized, and the Hund coupling 
$J_{\rm H}$ is so strong that the spin of the $e_{\rm g}$ electrons obey 
that of the $t_{\rm 2g}$ spins at each site.
Then the carrier becomes essentially spinless, and the hopping 
$t_{ij}$ between the two sites are determined by the relative relation 
of the two localized spins.
\begin{equation}
t_{ij} = t\left( 
\cos{ {\theta_i} \over 2 }\cos{ {\theta_j} \over 2 }
+ e^{ - {\rm i}(\phi_i - \phi_j)} 
\sin{ {\theta_i} \over 2 }\sin{ {\theta_j} \over 2 }
\right),
\end{equation}
where the direction of the localized spin at site $i$ is 
represented by the polar coordinates $\theta_i$ and $\phi_i$.
The amplitude $|t_{ij}|$ is given by 
$t \cos ( \theta_{ij} / 2) $ with $\theta_{ij}$ being the angle
between the two spins.\cite{REFanderson}
The phase of $t_{ij}$, on the other hand, constitutes the gauge field 
discussed intensively 
in the context of RVB theories.\cite{REFbaskaran,REFnagaosa}
In this paper we focus on the amplitude of $t_{ij}$ considering the
magnetic domain wall.
When the spins are inverted between two layers, i.e., $\theta_{ij}= \pi$,
The hopping $t_{ij} =0$ and the
conductance is zero through that domain wall.
However this idealistic situation is rather difficult to 
realize, because the width of the magnetic domain wall $L$ is determined by
the competition between the exchange interaction and the magnetic anisotropy 
energy. 
Recent neutron scattering experiment \cite{REFhirota}
has revealed the spin wave dispersion
in LaMnO$_3$, where the spin align ferromagnetically 
within the plane while antiferromagnetically ordered between planes.
They observed the gap in the spin wave dispersion,
from which they estimated the anisotropy energy 
$\Delta$ $=g \mu_{\rm B} H_{\rm A}$
$=0.61 \pm 0.11$ meV which is about 1/20 of the exchange coupling 
$8JS$ $=13.36 \pm 0.18$ meV within the layer.
A rough estimate of the domain wall width $L$ is given by 
$L \sim \sqrt{8JS/\Delta} \sim 5$
because $L$ is determined by the balance between the elastic energy 
$8JS \cdot L(1/L^2)$ and the anisotropy energy $\Delta \cdot L$.
We then have to consider the conductance through a finite-width domain wall.

\begin{figure}[b]
\centerline{\epsfile{file=dw.eps,width=80mm,height=15mm}}
\caption{ Spin configuration of a domain wall.}
\label{figdomainwall}
\end{figure}

In this paper we study the conductance through a magnetic domain wall 
as a function of its width $L$ and the energy of the incoming carrier.
The model we employ is the one-dimensional tight-binding Hamiltonian. 
\begin{equation}
H = - \sum_i t_i^{\phantom{\dagger}}
C^{\dagger}_{i} C_{i+1}^{\phantom{\dagger}} + h.c.
\label{eqhamil}
\end{equation}
where $t_i=t_{i,i+1}$ is given by 
$t_i =t \cos(\pi/2L)$ for $0 \le i \le L$ while $t_i=t$ for other $i$.
One may worry about the degeneracy of $e_{\rm g}$ orbitals, which is especially 
important when one consider the density of carriers.
The basic observation is that the carrier number is the density of the holes
doped to the Mott insulator. This is controlled by the Sr concentration $x$
in La$_{1-x}$Sr$_x$MnO$_3$. 
For $x=0$ the system is a Mott insulator in the sense that the 
$e_{\rm g}$ orbitals for each Mn ion is singly occupied.
This is because the strong on-site Coulomb interaction 
prohibits the double occupancy of $e_{\rm g}$ orbitals for each site. 
This constraint is treated by introducing the 
rotating coordinates with the isospin $\vec T_i$ for the orbital degrees of
freedom together with the usual spin $\vec S_i$.
The repulsive on-site interaction between the electrons in the 
$e_{\rm g}$ orbitals is treated by
introducing the Stratonovich-Hubbard variables 
$\vec \phi_{Ti}^{\phantom{\dagger}}$ and
$\vec \phi_{Si}^{\phantom{\dagger}}$ corresponding
to $\vec T_i$ and $\vec S_i$, respectively.
Then the  $e_{\rm g}$ electrons 
$d^{\dagger}_{i \gamma \alpha}$, 
$d_{i \gamma \alpha}^{\phantom{\dagger}}$
( $\alpha$: spin index, $\gamma$: orbital index )
are coupled with these fields as
$ \sum_{\alpha} {\vec \phi}_{Ti}^{\phantom{\dagger}} \cdot 
d^{\dagger}_{i \gamma \alpha}
{\vec \sigma_{\gamma \delta}}^{\phantom{\dagger}} 
d_{i \delta \alpha}^{\phantom{\dagger}}$ and  
$ \sum_{\gamma} {\vec \phi}_{Si}^{\phantom{\dagger}} \cdot 
d^{\dagger}_{i \gamma \alpha} {\vec \sigma_{\alpha \beta} }^{\phantom{\dagger}}
d_{i \gamma \beta}^{\phantom{\dagger}}$. In the rotating frame the 
transformed electrons are feeling
both the $\vec \phi_T^{\phantom{\dagger}}$
and $\vec \phi_S^{\phantom{\dagger}}$ fields
in the $+z$ direction, and hence the density of states are split
into four. The lowest one corresponds to $\uparrow$ both for the spin and 
isospin, which is fully occupied for $x=0$. 
The spin- and orbital-less operator $C^\dagger$, $C$ appearing in eq.(2)
is the hole operator in this lowest bunch of the density of states,
and their density is the Sr concentration $x$.
Then in reality the effective hopping integral $t_{ij}$ depends also on the 
isospin $\vec T$ as well as the spin $\vec S$.
For the magnetic domain wall we have to determine also the spatial
variation of the isospin. In this paper, however, we neglect this 
orbital degrees of freedom. We only mention that the conductance $G$ should be 
reduced if one consider this factor, and $G$ obtained in this
paper gives the upper bound.

We calculate the transmission coefficients of the system (\ref{eqhamil})
by using the transfer matrix method and the conductance
by the Landauer formula.
By using a one-particle state which is written as
$\vert \Psi \rangle 
= \sum_{i=1}^{N}
\psi_{i}^{\phantom{\dagger}} C_{i}^{\dagger}
\vert 0 \rangle$,
where $\psi_{i}$ are complex coefficients,
the corresponding Schr\"odinger equation,
$H \vert \Psi \rangle = E \vert \Psi \rangle$,
where $E$ is the energy eigenvalue, is
\begin{equation}
-t_{i+1}\psi_{i+1}
-t_{i-1}\psi_{i-1}
=E\psi_{i}.
\label{eqschrodinger}
\end{equation}
This equation can be written as
\begin{eqnarray}
\Psi_{i+1} =
M_i \Psi_i,
\end{eqnarray}
where
\begin{eqnarray}
\Psi_i =
\left(
 \begin{array}{c}
  \psi_{i} \\
  \psi_{i-1}
 \end{array} 
\right), \ \
M_i =
\left(
 \begin{array}{cc}
-\frac{E}{t_{i+1}} & -\frac{t_{i-1}}{t_{i+1}} \\
1 & 0
\end{array} 
\right).
\end{eqnarray}
These matrices describes the propagation of a plane wave
and their two eigenstates are corresponding
to the left- and right-going solutions, 
$\vec e_i(+)$ and $\vec e_i(-)$, respectively.
Any state can be represented by a linear combination of them as
\begin{eqnarray}
\psi_{i} = \left(\vec e_i(+), \vec e_i(-)\right)
\left(
 \begin{array}{c}
  c_1 \\
  c_2
 \end{array} 
\right)_i,
\end{eqnarray}
where $c_1$ and $c_2$ are the expansion coefficients.
When the domain wall locates between the $0$-th and $L$-th sites,
the transfer matrix 
\begin{eqnarray}
M = \prod_{i=-1}^{L} M_i, 
\end{eqnarray}
relates the incoming and outgoing amplitudes as
\begin{eqnarray}
\left(
 \begin{array}{c}
  t \\
  r'   
 \end{array} 
\right)_{L+1}
 = M
\left(
 \begin{array}{c}
  1 \\
  r
 \end{array} 
\right)_{-1},
\label{eqcoupledeq}
\end{eqnarray}
where $(t$, $r')$, and $(1$, $r)$, are the expansion coefficients
describing the wave amplitudes on the left and right
ferromagnetically ordered regions, respectively. 
To obtain the reflection and transmission amplitudes,
($r$ and $t$, respectively,) 
we solve the coupled equation (\ref{eqcoupledeq}) 
under the restriction $r'=0$.
By using the Landauer formula, the conductance is obtained as 
\begin{eqnarray}
G=\frac{e^2}{h}T,
\end{eqnarray}
where $T=t^* t$ is the transmission coefficient.
Since the analytical form as a function of $E$ and $L$ is
tedious to write down,
we only show the results in Fig.~\ref{figresult1},
where the conductance is shown as a function of the energy $E$
for $L=2$, 3, 5, and 10.

\begin{figure}[b]
\vspace{5mm}
\centerline{\epsfile{file=ccl.eps,width=85mm,height=60mm}}
\vspace{5mm}
\caption{ Conductance through a domain wall as a function of the 
energy $E$ for various domain wall width $L$.}
\label{figresult1}
\end{figure}

It is noted that the conductance $G$ is suppressed below a certain
energy $\epsilon_{\rm c}^{(L)}$ which depends on the domain wall width $L$.
This means that the magnetic domain wall acts as a energy filter.
This can be understood as follows.
In the domain wall region, the hopping is reduced 
from $t$ to $t \cos(\pi/2L)$.
Therefore if $L$ is large enough, 
we can consider the band structure in the domain wall region
($0 \le x \le L$) where the band width is reduced by the factor 
$\cos(\pi/2L)$.
The lower band edge is shifted upward to 
$\epsilon_{\rm min}^{(L)}$ $=-2t \cos(\pi/2L)$ for $0 \le x \le L$,
and the particle with energy less than $\epsilon_{\rm min}^{(L)}$
should tunnel though the potential barrier. 
Roughly $\epsilon_{c}^{(L)}$ coincides with $\epsilon_{\rm min}^{(L)}$.
When one consider the 1D Schr\"odinger equation 
\begin{eqnarray}
-\frac{\hbar^2}{2m} \frac{d^2 \psi}{dx^2} + V(x) \psi = \epsilon \psi
\end{eqnarray}
where $\frac{\hbar^2}{2m}=t$, $\epsilon=E+2t$, and the potential $V(x)$
being given by
\begin{eqnarray}
V(x)=
\left\{
 \begin{array}{cl}
        V=2t \left(1-\cos\frac{\pi}{2L}\right) &   \ \ \ 0<x<L \\
        0                       &   \ \ \ \mbox{otherwise}
 \end{array}
\right.
\end{eqnarray}
Then the conductance $G_{\rm s}(\epsilon)$ 
for this continuum model is obtained as
\begin{eqnarray}
G_{\rm s}(\epsilon)
=\frac{1}
{
1+\frac{V^2}{4\epsilon(V-\epsilon)}
\sinh^2 \left( L\sqrt{\frac{2m(V-\epsilon)}{\hbar^2}} \right)}.
\label{eqconticond}
\end{eqnarray}
From eq.(\ref{eqconticond}), it can be seen that $G_{\rm s}(\epsilon)$
$\sim \frac{e^2}{h}L^2 \epsilon$ for $\epsilon \ll t/L^2$ 
and
$G_{\rm s}(\epsilon)$ $\sim \frac{e^2}{h}$ for $\epsilon \gg t/L^2$.
This behavior is seen  in the large $L$ case, e.g., $L=10$,
in Fig. 2.

Now we discuss the possible application of our results to manganese oxides.
As discussed above the undoped LaMnO$_3$ shows ferromagnetism within 
the layer while the moments align antiferromagnetically 
between the layers. 
Then the dilute hole carriers doped to it should be able to move freely 
within the plane.
However the experiments show insulating behavior in 
La$_{1-x}$Sr$_x$MnO$_3$ for small $x$.\cite{REFtokura}
One possible explanation is that the local Jahn-Teller distortion
occurs around the carrier (Jahn-Teller polaron)
and finally leads to the self-trapping.\cite{REFMills,REFzhang}
Another possibility is that  each plane is divided into several 
magnetic domains, and the domain walls block the carriers.
When the carrier concentration is $x$,
the Fermi wavevector $k_{\rm F}$ is $(4\pi x)^{1/2}$ assuming that
the hopping between the layers is zero due 
to the antiferromagnetic ordering. 
Then the energy $\epsilon$ for the occupied states measured
from the bottom of the band satisfies
$\epsilon \le t k_{\rm F}^2= 4\pi t x$ which should be 
compared with $\epsilon_{\rm c}(L)$.
Assuming $L=5$, $\epsilon_{\rm c}(L) + 2t$ can be read to be around
$0.2t$ from Fig. 2.
Then for $x \le x_{\rm c} =\frac{0.2t}{4\pi t} \cong 0.02$, domain walls
block the carriers and the system will become
insulator.

In summary we have studied the effect of a magnetic domain wall
on the transport in the double exchange system. 
It is found that even a smoothly varying domain wall reflects the
carriers  when the carrier concentrations is low 
enough, and this might explain the insulating behavior
of La$_{1-x}$Sr$_x$MnO$_3$ for small $x$.
In 2 and 3D the curvature of the domain wall will
affect the conductance,
and its fluctuations generate the noise. 
This may explain the $1/f$ noise observed experimentally,
but the detailed analysis in now in progress.
We also propose that this sensitivity of the conductance 
to the magnetic domain wall in slightly doped Mott insulator
will give a unique opportunity for the device made of strongly correlated 
electronic systems.

\acknowledgements
We are grateful to Y. Tokura, Y. Endoh, and S. Ishihara 
for useful discussions. This work is supported by Grant-in-Aid for 
Scientific Research No. 04240103
from the Ministry of Education, Science, and Culture of Japan.
One of the authors (M.Y.) acknowledges financial support
from the JSPS Research Fellowships for Young Scientists.

\bibliographystyle{prsty}

\end{document}